\documentclass[prl,aps,preprint,nofootinbib,showpacs,floatfix]{revtex4}
\usepackage{graphicx,amssymb,amsmath}
\usepackage{epsf}
\usepackage{pslatex}
\usepackage{dcolumn}
\begin{document}
\bibliographystyle{apsrev}
\epsfclipon


\newcommand{\pbp}{\langle \bar \psi \psi \rangle}
\newcommand{\pbdmdup}{\left\langle \bar \psi \frac{dM}{du_0} \psi
\right\rangle}
\newcommand{\bea}{\begin{eqnarray}}
\newcommand{\eea}{\end{eqnarray}}
\newcommand{\be}{\begin{equation}}
\newcommand{\ee}{\end{equation}}
\newcommand{\nn}{\nonumber}
\newcommand{\VEV}[1]{\left\langle #1\right\rangle}
\def\maths#1{$#1$}
\newcommand{\Tr}{\mbox{Tr}}
\def\lsim{\raise0.3ex\hbox{$<$\kern-0.75em\raise-1.1ex\hbox{$\sim$}}}


\title{Quark-Qluon Plasma in an External Magnetic Field}

\author{L. Levkova$^1$ and C. DeTar$^1$} \affiliation{$^1$Department of Physics and Astronomy, University of Utah,
Salt Lake City, Utah 84112, USA}

\begin{abstract}
Using numerical simulations of lattice QCD we calculate the effect of an external magnetic field on the equation of state  
of the quark-gluon plasma. The results are obtained using a Taylor expansion of the pressure with respect to the
magnetic field for the first time. 
The coefficients of the expansion are computed to second order in the magnetic field. Our setup for the external magnetic field
avoids complications arising from toroidal boundary conditions, making a Taylor series expansion straightforward. This study is exploratory and is meant to serve as a proof of principle.
\end{abstract}
\maketitle

\newpage

\section{Introduction}

The behavior of the quark-gluon plasma in the presence of a strong magnetic field
is of interest to cosmology, astrophysics, and heavy-ion collisions.
Shortly after the big bang, when one of the main components of the Universe was the quark-gluon plasma,
strong magnetic fields of $O(10^{16}\, {\rm T})$ and higher may have existed as a result of 
the nonequilibrium dynamics of the electroweak phase transition, generation of topological defects and other phenomena. 
(For a review of some of these mechanisms see Ref.~\cite{Grasso:2000wj}.) This could have affected the
subsequent structure formation and evolution of the Universe, since the equation of state of the plasma 
could have been modified by these fields. Very strong magnetic fields are also generated   
in the vicinity of magnetars [$O(10^{11}\, {\rm T})$] and if these fields permeate the interior of such a star,
they may affect the state of the high-density hadronic matter in its core and thus potentially influence the star's properties such as
its temperature and diameter-to-mass ratio \cite{magnetars}.
Currently, the properties of the quark-gluon plasma are studied
at the LHC, RHIC, and other experimental facilities, and its equation of state is important in the
process of predicting the features of the particle spectra created in a heavy-ion collision. 
In a noncentral heavy-ion collision strong magnetic fields are induced by the spectator protons 
in the nuclei moving with speeds close to the speed of light \cite{hic}. (There is a smaller contribution 
from the participant region.) The almond-shaped volume of the
developing quark-gluon plasma 
is immersed in this external magnetic field, which is estimated to be of $O(10^{15}\, {\rm T})$. 
It appears that if such a strong magnetic field modifies the properties of the plasma, the particle
spectra produced might also be affected.   

In this Letter we attempt to calculate the effect of a strong external magnetic field on the pressure of the quark-gluon plasma.
We use a Taylor expansion method and calculate the contribution to the pressure up to a second order in the field. At lower
temperatures we compare with the pressure calculated using the hadron resonance gas (HRG) model \cite{Endrodi:2013cs}.
This Letter is organized as follows: 
The first section presents the particulars of introducing
an external magnetic field on a torus, both in the continuum and on
a discrete lattice. We give our preferred way of dealing with them. In the next section, the Taylor expansion method for the pressure is    
described. The final section gives our results and conclusions.   
 
\section{Magnetic field on a torus}

The introduction of a constant magnetic field on a (continuum space) torus leads to peculiar requirements
such as the quantization of the magnetic flux and breaking of translational invariance to a
discrete group \cite{AlHashimi:2008hr}. In other words, if the torus 
is of size  $L_x\times L_y$ then the magnetic field $B$ in the $\hat z$ direction should have the magnitude
$ B=(2\pi b)/(|q|L_x L_y)$, $b\in \mathbb{Z}$.
This relation follows from the requirement for (1) gauge invariance of the wave function of a particle with charge of magnitude $|q|$
under shifts of size $L_x$ and $L_y$ and (2) the periodic boundary conditions in both
the $\hat x$ and $\hat y$ directions. In Ref.~\cite{AlHashimi:2008hr} it is also shown that the Polyakov 
loops are not translationally invariant in the $\hat x$ and $\hat y$ directions, unless the translation 
is done by shifts which are integer multiples      
of $a_x=L_x/b$ and $a_y=L_y/b$.

The lattice representation of space-time is usually a discretized torus; thus, all of the quantization 
rules described above are applicable in this case, albeit in their even more restrictive discretized version.
It follows that the magnetic field on the lattice (choosing $B$ in the $\hat z$ direction again) is quantized as:
$|e|B=(6\pi b a^{-2})/(L_xL_y)$, $0<b<L_xL_y/2$,
 where $b$ is an integer. The meaning of $L_x$ and $L_y$ is changed to be the number of lattice
points in the $\hat x$ and $\hat y$ directions, and $a$ is the lattice spacing. The additional factor of $3$ in the numerator  
originates from the fractional quark charges, the smallest of which is $-|e|/3$ and thus determines
the quantum of the magnetic field. The maximum value of the integer $b=L_xL_y$, due to the finite dimensions of the
lattice, is divided by a factor of 2, since measurements on the lattice  will typically show a  
symmetry with respect to the mid-lattice points, and this further restricts the number of physically different values for 
$b$.   
If $B$ is not quantized, at minimum there is one corner plaquette where 
the magnetic flux through it is equal in magnitude to the magnetic flux through the rest of the 
lattice and opposite in direction: $\Phi_{\rm corner}=-B(L_xL_y-1)a^2$, modulo the magnetic quantum. This is natural since
without quantization, the net flux through the $\hat x$-$\hat y$ surface must be zero ({\it i.e.},
the flux lines going into the surface inevitably have to reappear out of the surface at some corner of the
lattice). 

The magnetic field setup described above is not well suited for a calculation
using a Taylor expansion method of a bulk thermodynamic quantity.
In order to obtain the Taylor expansion coefficients, we need to take
derivatives with respect to the magnetic field. But taking a derivative  
with respect to a quantized quantity (in this case the magnetic field) is not 
straightforward to implement or interpret. If we decide to ignore the quantization
of $B$ and treat it as a continuum variable, the corner plaquette
with a large magnetic flux dominates the bulk observables and completely
skews the physics. This happens because bulk observables are sums over the
whole lattice volume, and the corner plaquette cannot be simply excluded from that sum.
(There are cases where the corner plaquette ``defect'' is of less importance, such as when 
particle propagators are calculated at a ``safe'' distance away from it, since they may not involve 
a sum over the whole lattice \cite{Rubinstein:1995hc}.) 

These difficulties prompted us to change the magnetic
field configuration from the above setup to one where the magnetic field is in
the $\hat z$ direction on one half of the lattice and in the $-\hat z$ direction on the other half (for brevity
we call it the ``half-and-half setup''). The obvious advantage 
is that we don't need to quantize the magnetic field, because the flux from one half     
of the lattice comes out from the other without being large for any size of closed
loop on the lattice. Thus the application of the Taylor expansion method becomes straightforward for 
thermodynamic quantities. In addition, with our method the pressure is isotropic, since derivatives of 
the partition function are taken at a constant, nonquantized external magnetic field (for a discussion of
pressure anisotropies 
see Ref.~\cite{Bali:2013esa}).  

We do not expect that the thermodynamics of the system is much affected
by the fact that the magnetic field changes direction in the middle and the end of the lattice, as long as the
spatial volume is suitably large. Generally, the pressure and energy density should not depend on 
the field direction, and if the two lattice halves are thermodynamically large, the surface defects introduced
in the middle and the end of the lattice should have a small effect on the final results. 
But of course for finite lattices, the effective halving of
the spatial volume may lead to increased finite volume effects [probably of $O(1/L_s)$], which should be estimated by 
comparing results on different lattice volumes.  

The realization of the half-and-half setup we work with has the $U(1)_{\rm EM}$ links as:
$u_{\hat y}(B,q,X) = e^{ia^2qBx^\prime}$, with
$x^\prime = x-L_x/4$ for $x\leq L_x/2$ and $x^\prime = 3L_x/4- x$ for $x> L_x/2$; $u_{\hat x,\hat z,\hat t}(B,q,X)=1$.
This choice defines particular values of the
link phases ({\it i.e.}, the vector potential values) symmetric with respect to the mid-lattice points. 
In fact, any choice of the phases such that they increase by
an additional $ia^2qB$ in the $\hat x$ direction at each lattice point on half of the lattice and decrease by 
that amount on the other half, constitutes a valid half-and-half magnetic field setup.
However, the different choices are not gauge-equivalent. 
They give different values of Polyakov loops in the $\hat y$ direction,
leading to small differences in physical observables; but those differences should vanish at
infinite volume.
The phase choice 
has proven to have a large effect on the stochastic noise in the measured observables.
The Taylor expansion coefficients of the pressure can be thought of as sums of $n$-point lattice loops
with insertions (multiplication) of the phases. 
For a particular loop the sum over all such insertions is gauge invariant, and therefore requires cancellations
among the noisy gauge-specific terms, when they are estimated using random vectors.
Since the subtraction between the terms
in the loop is stochastic, the smaller the  magnitudes of 
these link phases, the less noise is introduced in the measurement.

It can be easily seen that the phase configuration we work with (defined up to a constant lattice translation) 
fulfills all 
the above conditions ({\it i.e.}, both symmetry and small magnitude of the phases), and minimizes the noise.
This ``minimal-noise'' choice of the vector potential on the lattice,
decreases the standard deviation by a sizable factor in the measurement of the
second order Taylor coefficient in comparison with other choices. 
 
\section{Taylor expansion of the pressure with respect to the magnetic field}

As explained in the previous section, the half-and-half configuration for the magnetic field 
frees us from the necessity to quantize the magnetic field. Thus a Taylor expansion
of various thermodynamic variables is straightforward. The direction of the magnetic
field should not affect the pressure, which is a $CP$ invariant. Moreover, the setup 
of the magnetic field that we use is explicitly $CP$ invariant, so we know that
the odd-order Taylor expansion coefficients of the pressure are zero. (But a $CP$-even result also should be expected if 
we used a magnetic field configuration, where the field is quantized and points only in one direction.)        

To calculate the Taylor expansion coefficients of the pressure, we chose a method similar to the one used to
obtain them in the case of an expansion with respect to a nonzero chemical potential \cite{DeTar:2010xm}.
One crucial difference from the nonzero chemical potential case is that here we may need 
to renormalize some of the Taylor expansion coefficients . This is due to  
the UV divergence in the QCD+QED theory which is of second order in the external 
magnetic field and which in principle can be
absorbed in the electric charge renormalization. (See Ref.~\cite{Endrodi:2013cs} for a discussion
in the context of the HRG model.) On the lattice, unless the
operators we work with are somehow renormalized and have no $O(B^2)$ UV divergence by construction, we need to cancel 
this divergence by subtracting, for example, an appropriate zero-temperature correction. 
In the Taylor expansion method for the pressure, only the second order coefficient contains the $O(B^2)$ UV divergence; 
hence, it is the only term
in need of a zero temperature correction. Still, this increases the
computational cost. On the other hand, generating zero- and nonzero-temperature gauge ensembles with explicit background 
magnetic fields \cite{Bonati:2013lca} is generally costlier (at least when compared with a calculation up to the second order in the Taylor expansion), 
while here we can use preexisting ensembles generated with $B=0$. 
    
Another important difference with the nonzero chemical potential case is that the QCD vacuum 
is modified by the presence of the magnetic field (but not of the chemical potential) and the pressure is nonzero even at zero temperature.
This vacuum pressure is a nonthermal contribution to the whole pressure, and its lowest order 
is $O(B^4)$. Hence, the $O(B^2)$ coefficient in the expansion of the pressure (after renormalization) should be zero
at zero temperature, while the higher order coefficients will have nonzero values. In other words, the second order coefficient
 has an entirely ``thermal'' origin and will be nonzero only if the temperature is nonzero. On the other hand, the 
fourth and higher order coefficients have both vacuum 
and thermal contributions.  

In the case 
of 2+1 staggered-type quark flavors, the Taylor expansion of the pressure $p$  at temperature $T$ with respect to the dimensionless parameter $|e|B/T^2$ is shown below
(before renormalization):
\be
\frac{p(T)}{T^4}=\frac{\ln Z(B)}{T^3V} = \sum_{n=0}^{\infty}C_{n}(T)(|e|B/T^2)^{n},\hspace{1cm}
C_{n}(T)=\frac{L_t^3}{L_s^3}\frac{1}{n!}\left.\frac{\partial^{n} \ln Z(B)}{\partial (|e|B/T^2)^{n}}\right|_{B=0},
\ee
where the partition function is
$Z(B) = \int dU e^{-S_g} e^{U}\, e^{D}\,e^{S}$,
with $U=\ln\det M^u(q_u,B)/4$, $D=\ln\det M^d(q_d,B)/4$, $S=\ln\det M^s(q_s,B)/4$, and $S_g$ is the
gluon action. The charges of the up, down, and the strange quarks are denoted as $q_u, q_d$, and $q_s$.
The spatial dimension of the lattice is $L_s(=L_x=L_y=L_z)$ and the temporal one is $L_t$.
For simplicity we will set the magnitude of the electron charge $|e|=1$ and thus omit it in the following expressions.
In the half-and-half setup for the magnetic field in the $\hat z$ direction, the fermion matrix for 
a given quark flavor $f=u,d,s$ is:
$M_{X,Y}^f(B,q_f) = am_f\delta_{X,Y}+ D^{\hat z,\hat t,\hat x}_{X,Y} + D^{\hat y}_{X,Y}(B,q_f)$,
where $m_f$ is the quark mass for flavor $f$ and $D^{\hat z,\hat t,\hat x}_{X,Y}$ is a sum of the  Dirac 
operators in the $\hat x$, $\hat z$, and $\hat t$ directions at all points.
This term does not explicitly depend on $B$. The dependence on $B$ is in the third 
term only. 
 As noted in the previous section, we work with the half-and-half magnetic field configuration, which ensures
that the stochastic noise in the measured observables is minimized.
 As in Ref.~\cite{DeTar:2010xm}, it is convenient to define the observable:
\be
A_{nml}=\frac{1}{q_u^nq_d^mq_s^l}\VEV{e^{-U}e^{-D}e^{-S}\frac{\partial^ne^{U}}{\partial (a^2B)^n}\frac{\partial^me^{D}}
{\partial (a^2B)^m}\frac{\partial^le^{S}}{\partial (a^2B)^l}}.
\label{eq:As}
\ee
We work with equal $u$ and $d$ quark masses, which means that the following symmetry holds:
$A_{nml}=A_{mnl}$. 
It is straightforward to show that the first two coefficients 
in the pressure expansion [after $C_0(T)$, which is the pressure at $B=0$] are:
\bea
\label{eq:C1}
C_1(T) &=& \frac{L_t}{L_s^3}\left[(q_u+q_d)A_{100} + q_sA_{001}\right],\\
C_2(T) &=& \frac{1}{2L_tL_s^3}\left[(q_u^2+q_d^2)A_{200} + q_s^2A_{002} +
2q_uq_dA_{110} + 2(q_u+q_d)q_sA_{101}-\left(\frac{L_s^3}{L_t}C_1\right)^2\right].
\label{eq:C2}
\eea
The explicit forms of the $A_{nml}$'s used in the above are easy to obtain from 
Eq.~(\ref{eq:As}).
We calculate the necessary $A_{nml}$'s stochastically using a number of random Gaussian sources 
for each lattice configuration. For more details, such as the parameters of the lattice ensembles and 
number of sources, see Table~I.   
\begin{table}[t]
\begin{small}
\begin{tabular}{|c|c|c|c|c|c|r|r|r@{.}l|}
\hline\hline
$T$ [MeV] & $\beta$ & $m_l/m_s$ & $V_{T\neq0}$ & $V_{T=0}$ & \#RS$_{T\neq0}$ & \#RS$_{T=0}$ & $C_1^r\times 10^{-4}$ & \multicolumn{2}{c|}{$C_2^r\times 10^{-3}$}  \\
\hline
134 & 6.195 & 0.00440/0.0880 & $32^3\times8$ & $32^3\times32$ & 2400 & 400 & $1(5)$& $-0$&$3(5)$\\
154 & 6.341 & 0.00370/0.0740 & $32^3\times8$ & $32^3\times32$ & 2400 & 500 & $6(5)$ &  0&4(4)\\
167 & 6.423 & 0.00335/0.0670 & $32^3\times8$ & $32^3\times32$ & 1200 & 200 & $-8(4)$& 2&18(53)\\
167 & 6.423 & 0.00335/0.0670 & $48^3\times8$ & $48^3\times48$ & 1200 & 400 & $-4(3)$& 2&28(51)\\
173& 6.460 & 0.00320/0.0640 &  $32^3\times8$ & $32^3\times64$ & 1200 & 200& $-8(6)$ & 3&81(55)\\
227& 6.740 & 0.00238/0.0476 &  $32^3\times8$ & $48^3\times48$ &1200 & 200&  $-1(2)$ & 10&4(0.8) \\
373 & 7.280 & 0.00142/0.0284 & $32^3\times8$ & $48^3\times64$ &1200 & 40 &   $1(1)$ & 19&3(1.3)\\ 
\hline\hline
\end{tabular}
\end{small}
\caption{Parameters of the ensembles in this study and the number of Gaussian random sources (\#RS) used on each configuration
at zero and nonzero temperature $T$.
The number of gauge configurations is $50$ for each ensemble and temperature, except for the $\beta=6.46$ and
 6.423 (for the larger volume) , where
they are 60 and 70 at $T=0$, respectively.} 
\label{tab:param}
\end{table}
As explained at the beginning of the section, we need to subtract the zero-temperature corrections from the
second order Taylor coefficient. Thus we have:
$C_1^r(T) =  C_1(T)$ and 
$C_2^r(T)=  C_2(T) - C_2(0)$,
where the superscript $r$ denotes a renormalized observable.

\section{Results and conclusions}

For this exploratory work we employ (2+1)-flavor lattice ensembles generated by the HotQCD Collaboration using 
the HISQ/tree action (at $B=0$) along a line of constant physics with $m_l/m_s=0.05$ \cite{hisq},
the parameters of which are given
in Table~I. The high-temperature ensembles have $L_t=8$ and encompass temperatures between 134 and 373 MeV (the temperature scale is
as in Ref.~\cite{hisq}).
To calculate the necessary observables from Eqs.~(\ref{eq:C1}) and (\ref{eq:C2}), we use Gaussian random sources, the number of which is
also given in Table~I. The computational cost for the whole calculation is around 
30~000 GPU-hours using QUDA \cite{GPU}.

In the last two columns of Table~I, the results for $C^r_1$ and $C^r_2$ are presented. As expected,
$C^r_1$ is compatible with zero at any temperature. Hence, the first nontrivial contribution to the pressure
is coming from the second order coefficient $C^r_2$, which can be interpreted as the magnetic
susceptibility of the quark-gluon plasma. In the left panel of Fig.~1 we examine its behavior:
it shows an increase with $T$ in the temperature region studied.
The magnetic susceptibility $C^r_2$ is positive for temperatures above the transition,
which means that the quark-gluon plasma exhibits a paramagnetic behavior to lowest (linear) order in the magnetic field.
Paramagnetism in the quark-gluon plasma has been also previously found in  other lattice studies~\cite{Bali:2013esa,Bonati:2013lca}.

The values of $C^r_2$ at the two lowest available temperatures (134 and 154 MeV) are compatible with zero,
and clearly require more statistics for their determination. Moreover, these points are most likely 
to be affected by the lattice cutoff since they have coarser lattice spacings. We also tried to estimate the finite volume effects 
by recalculating $C^r_2$ at $T=167$ MeV on a larger spatial volume of $48^3$ for both the zero- and nonzero-temperature ensembles.
The red empty circle denotes this result, and, as one can see, the finite volume effects are entirely within the
statistical error of the calculation, since the small and large volume values are compatible.
\begin{figure}[t]
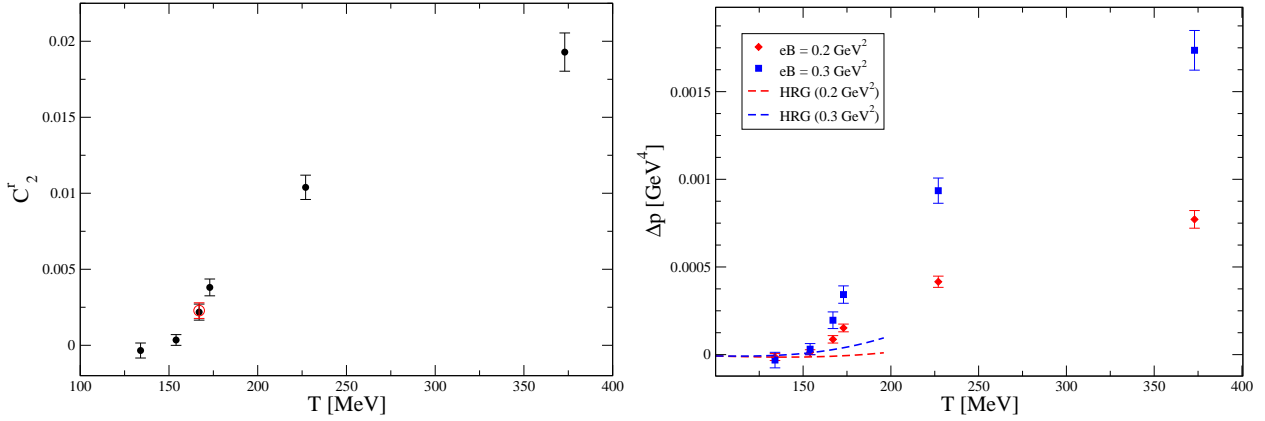

\begin{center}
\begin{tabular}{cc}
\includegraphics[width=0.5\textwidth]{C2r.eps}& \includegraphics[width=0.5\textwidth]{dp_thermal.eps}
\end{tabular}
\caption{(Left panel) The second order coefficient in the Taylor expansion of $p/T^4$ {\it vs.} the temperature. The red empty circle 
at $T=167$ MeV denotes
a value calculated at a larger spatial volume in order to check for finite volume effects.  
(Right panel) The $O(B^2)$ thermal contribution to the pressure of the quark-gluon plasma due to the presence of a magnetic field, 
for two values $eB= 0.2$ and 0.3 GeV$^2$. The HRG results are from Ref.~\cite{Endrodi:2013cs}.
}
\label{fig:C2r}
\end{center}
\end{figure}

In the right panel of Fig.~1 we show the thermal contribution to the pressure due to the presence of
an external magnetic field to second order in the field i.e., without the vacuum pressure: $\Delta p = C^r_2 (eB)^2$. We show
$\Delta p$ for two values: $eB= 0.2$ and 0.3 GeV$^2$. 
We compare it with predictions of the   
HRG model \cite{Endrodi:2013cs} for the above fields (also with the vacuum pressure subtracted).
Up to about 154 MeV the HRG and our result are roughly compatible (within large errors),  and further studies 
should refine this statement.   
Of course, the HRG values contain terms of
all possible orders,
unlike our second-order-only result. At higher temperatures, the HRG and our results deviate, which is not surprising.

In light of our findings, it seems that the thermal correction of the pressure due 
to the presence of a magnetic field is small (within a few percent) for fields of $O(10^{15}{\rm T})$ [or equivalently $O(10^{-1}\rm GeV^2 )$],
which are relevant for heavy-ion collision experiments. 
On the other hand, if in the early Universe fields reached  $O(10^{16}{\rm T})$ [or $O(1\,\rm GeV^2 )$], this correction 
in the studied temperature range becomes large: roughly $20\% - 100\%$ (with the larger value corresponding to lower temperatures).
Of course, the latter statement assumes that using only the second order coefficient is sufficient to estimate $\Delta p$ in the presence of
such a large external magnetic field.  
 
In conclusion, in this work we argue for the feasibility and convenience of the Taylor expansion method for calculating the
equation of state of the quark-gluon plasma in the presence of an external magnetic field.
This work should be expanded in the future to include larger statistics,
study of the finite volume effects at different temperatures, and contributions of higher orders in the Taylor expansion of the pressure. Other quantities
such as the trace anomaly, energy density, effects on the chiral condensate, {\it etc}., can also be studied in this way 
with suitably higher statistics. (Results for some of these quantities obtained with other methods can be found in Refs.~\cite{Bali:2013esa,Bonati:2013lca,others}.)       

\begin{center} {\bf Acknowledgements}\end{center} 

We thank Urs Heller and Gergely Endr\"odi for helpful discussions.
Computations for this work were carried out with resources provided by the USQCD
Collaboration, the University of Utah Center for High Performance Computing, and Indiana University. This work
was supported in part by the U.S National Science Foundation under Grant No. PHY10-67881 and the U.S. Department of Energy under Grant No. DE-FC02-12ER-41879.

\end{document}